\documentclass[12pt,a4paper,vietnamese,british]{article}
\PassOptionsToPackage{natbib=true}{biblatex}
\usepackage[T5,T1]{fontenc}
\usepackage[utf8]{inputenc}
\usepackage{color}
\usepackage{babel}
\usepackage{array}
\usepackage{booktabs}
\usepackage[bookmarks=true,bookmarksnumbered=false,bookmarksopen=false,
 breaklinks=true,pdfborder={0 0 0},pdfborderstyle={},backref=false,colorlinks=true]
 {hyperref}
\hypersetup{pdftitle={What is ‘undone computer science’?},
 pdfauthor={Chantal Enguehard, Guillaume Munch-Maccagnoni and Alberto Naibo},
 linkcolor={dark-red},citecolor={dark-blue},urlcolor={blue},pdfborder={0 0 0},pdfborderstyle={}}

\makeatletter

\newcommand*\LyXThinSpace{\,\hspace{0pt}}
\pdfpageheight\paperheight
\pdfpagewidth\paperwidth

\providecommand{\tabularnewline}{\\}

\@ifundefined{date}{}{\date{}}
\usepackage{mlmodern}
\usepackage[p,scaled=0.95]{zlmtt}

\usepackage{csquotes}

\AtBeginDocument{
\renewbibmacro*{volume+number+eid}{%
  \printfield{volume}%
  \printfield{number}%
  \setunit{\addcomma\space}%
  \printfield{eid}}
\DeclareFieldFormat[article]{number}{\mkbibparens{#1}}
\AtEveryBibitem{\clearlist{language}}
\AtBeginBibliography{\sloppy}
\DefineBibliographyExtras{english}{\restorecommand\mkbibnamefamily}
\DefineBibliographyExtras{english}{%
  }
\setlength\bibitemsep{0.5\baselineskip}

\DeclareCiteCommand{\citeyear}
    {}
    {\bibhyperref{\printdate}}
    {\multicitedelim}
    {}
\DeclareCiteCommand{\citeyearpar}
    {}
    {\mkbibparens{\bibhyperref{\printdate}}}
    {\multicitedelim}
    {}

\DeclareDelimFormat[textcite]{nameyeardelim}{\addnbspace}
\defcounter{abbrvpenalty}{9000}
}

\usepackage{xcolor}
\usepackage{enumitem}
\usepackage{typearea}
\areaset{13,9cm}{23,0cm}

\usepackage[expansion=true,stretch=14]{microtype}

\usepackage{setspace}

  \definecolor{dark-red}{rgb}{0.6,0,0}
  \definecolor{dark-blue}{rgb}{0,0.0,0.6}
  \definecolor{blue}{rgb}{0,0,0.8}

\newcommand*{\doi}[1]{\textsc{doi:} \href{https://doi.org/#1}{\texttt{#1}}}

\usepackage[skip=0.2\baselineskip plus 2pt,indent,parfill]{parskip}

\makeatother

\usepackage[style=ext-authoryear,style=ext-authoryear, natbib=true, backend=biber, date=year, maxcitenames=3, mincitenames=2, maxbibnames=10, minbibnames=10, uniquename=init, language=autobib, autolang=hyphen]{biblatex}
\begin{document}
\title{\textbf{What is `undone computer science'?}}
\author{Chantal Enguehard\thanks{Nantes Université, LS2N, Nantes, France}\\
Guillaume Munch-Maccagnoni\thanks{INRIA, LS2N CNRS, Nantes, France}\\
Alberto Naibo\thanks{Université Paris 1 Panthéon-Sorbonne, IHPST, UMR 8590, Paris, France}}
\author{%
\begin{tabular*}{1\columnwidth}{@{\extracolsep{\fill}}r>{\raggedright}p{0.5\columnwidth}}
{\large\emph{Chantal Enguehard}} & {\small Nantes Université, LS2N, France}\tabularnewline\addlinespace[2ex]
{\large\emph{\hspace{-0.3em}Guillaume Munch-Maccagnoni}} & {\small INRIA, LS2N CNRS, Nantes, France}\tabularnewline\addlinespace[2ex]
{\large\emph{Alberto Naibo}} & {\small Université Paris 1 Panthéon-Sorbonne,\vspace{-0.5ex}
}\\
{\small IHPST, UMR 8590, France}\tabularnewline\addlinespace[2ex]
\end{tabular*}}

\maketitle
\noindent\textbf{Abstract:} The concept of ‘undone science’ emerged
in the 2010s in research in social sciences at the intersection of
studies on social movements and of science and technology studies.
It refers to research questions that are neglected, ignored, or left
unfunded, even though they deserve to be explored. The aim of this
special issue is to apply this concept to computer science, by examining
whether the way this discipline is structured (including its sociological,
economic, and political dimensions), as well as the paradigms that
shape it, make it possible to identify epistemological and ethical
questions that are crucial for its development and conception.\bigskip{}

{\renewcommand{\footnoterule}{\rule{0cm}{0cm}\vspace{0cm}}
\renewcommand{\thefootnote}{\fnsymbol{footnote}}
\footnotemark[0]\footnotetext[0] {\begin{flushright}Translation of `Qu'est-ce que la science informatique non faite\,?'
In:\\
\emph{Philosophia Scientiæ}, 30(2): \emph{Undone Computer Science}.
Chantal Enguehard, Guillaume Munch-Maccagnoni and Alberto Naibo (eds.),
June 2026, pp.~5–16, \doi{10.4000/16952}.\\
Translated by Richard Dickinson (Inist-CNRS).\nocite{EnguehardMMNaibo2026issue,EnguehardMMNaibo2026questce}\end{flushright}}}
\renewcommand{\thefootnote}{\arabic{footnote}}

\section{Introduction. From undone science to\protect \\
undone computer science}

The expression ‘undone computer science’ selected for the title of
this special issue may seem somewhat unusual and does indeed bring
up a number of questions such as: Where does it come from? What does
it actually designate? Why should this be the object of epistemological
thought? This introduction to the special issue aims to provide some
answers to questions of this kind and discuss the emergence of a broader
debate about the status and specificity of computer science as a scientific
discipline.

The idea of ‘undone computer science’ stems from the attempt made
to apply the concept of ‘undone science’ to the field of computer
science. According to \citet{Hess2016}, this concept has emerged
over the last fifteen years in social science research carried out
at the intersection of social movement studies and science and technology
studies.

The starting point for this research was the observation that the
dynamics of scientific production cannot be entirely explained at
the microsocial level, or by applying an internal perspective specific
to each scientific community (for example, by studying interactions
within research laboratories\footnote{\emph{Cf.} what is sometimes referred to as the ``sociology of laboratory
life''~\citep[see][]{Latour1979,Woolgar1984}.} and considering the scientific community to be relatively autonomous
from a social standpoint). When taking these dynamics into account,
it is also necessary to consider factors that are external to scientific
institutions and to adopt a macrosociological approach in which political,
economic and industrial dimensions are integrated to contribute to
the identification, development and structuring of knowledge.

And yet, certain social movements can give rise to forms of protest
with epistemic implications when they are in opposition to decisions
taken by political institutions or the agendas of industrial stakeholders.
For example, movements of this kind may challenge the distribution
or marketing of certain products or technologies by calling for more
detailed study of their effects, the aim being, in this case, to demonstrate
that they could be dangerous (as in the case of campaigns against
GMOs or the use of certain products in the agri-food industry). In
this way, some movements actually go so far as to posit alternative
scientific or technological approaches to those adopted by political
institutions or industrial actors, particularly in the framework of
ecological transitions.

In this context, the term ‘undone science’ refers to ``areas of research
identified by social movements and other civil society organizations
as having potentially broad social benefits that are left unfunded,
incomplete, or generally ignored.''~\citep[p. 445.]{Frickel2009}
Indeed, studies devoted to ‘undone science’ aim to understand the
emergence of potential subjects for scientific study, which can provide
responses to certain objectives or values which may be ethical, social
or health-related, for example, but remain overlooked for various
reasons.

This may be because the subjects have yet to be clearly identified
before protest movements draw attention to them.\footnote{Conversely, there may also be attempts by certain protest movements
to prevent certain research from being carried out.} Another possibility is that such subjects may be identifiable and
have already been taken up by certain researchers, but have been deliberately
marginalised by political, economic or industrial institutions, notably
through a lack of funding or recognition~\citep[pp. 447--448]{Frickel2009}. 

However, one particularly important aspect of the concept of ‘undone
science’ is its capacity to enable us to address several crucial philosophy
of science questions. These are epistemological questions like the
relationship between ignorance and knowledge and the transition from
one to the other and also questions linked to scientific practice,
such as the distinction to be made between what is recognised as being
scientific and what is not, or the relationship between scientific
research and value systems. In this way, the notion of ‘undone science’
appears as a prism through which certain classic themes of the philosophy
of science can be re-examined, one example being Kuhn’s theme of research
paradigms \citep{Kuhn1970}.\footnote{The reference to Kuhn is, moreover, explicit in \citet{Frickel2009}
and in \citet{Hess2016}.}

This issue introduces the task of engaging in a somewhat different
and, in some ways, more specific exercise, namely to apply the concept
of ‘undone science’ to a particular discipline\LyXThinSpace –\LyXThinSpace computer
science. The aim is to highlight epistemological and ethical thought
about this field, and, in this way, potentially bring to light sociological,
political, anthropological, economic, and other such forms of thought.
In reality, it is not entirely surprising that one might wish to apply
the concept of ‘undone science’ to computer science because, in the
space of just a few decades, the products of computer science have
become deeply embedded in many aspects of our everyday life\LyXThinSpace –\LyXThinSpace in
technology, work, social and institutional relations, media, culture,
and so on. Although computer science is a relatively young discipline,
it has thus come to encompass economic, social and political dimensions
over time. It is therefore natural that one may expect to find dynamics
typical of undone science within this field, particularly in terms
of industrial, economic and institutional biases in their different
and diverse forms.

Of course, the role played by economic interests is a focus in the
field of undone science given that this concept has been mobilised
in the context of health and the environment~\citep{Hess2016}. Indeed,
\citet{Abdalla2021} have stated that the actions of ‘Big Tech’\LyXThinSpace –\LyXThinSpace the
major global technology companies\LyXThinSpace –\LyXThinSpace are
reminiscent of the tactics deployed by ‘Big Tobacco’\LyXThinSpace –\LyXThinSpace the
major global tobacco companies\LyXThinSpace –\LyXThinSpace from the
1950s onwards to manipulate public opinion and academic discourse.
Such actions have been analysed in terms of ‘undone science’ as capable
of producing an ‘epistemic rift’, in which the flow of scientific
knowledge into the political sphere no longer occurs~\citep[pp. 55--56]{Hess2016}.
For example, this kind of process is reportedly at work within ‘tech
ethics’, which is said to be ``absorbed by corporate logic and incentives'',
and serves as a form of ‘ethics-washing’ for the technology industry
in Green’s view~\citeyearpar{Green2021}.

One strong point of the concept of ‘undone science’ is that it can
be applied in areas beyond extreme situations of this kind. For example,
\citet{Hooker2021} describes an industrial bias, which leads to certain
research projects becoming somewhat like ``lotteries'' because of
the existing hardware and software constraints at the time of the
experiments involved. Constraints of this kind have sometimes played
a decisive role in determining which ideas have ‘won’ and which ‘lost’.
Instrumentation bias is a classic topic in the Kuhnian tradition,
but Hooker more specifically describes the industrial conditions that
mean hardware development tends to favour existing commercial applications
over serving as a testing ground for research ideas.\footnote{For example, graphical processing units (GPUs) that were developed
over a long period for the video game and computer graphics markets
before their actual usefulness was demonstrated for a broader range
of applications involving inherently parallel computations and including
deep learning~\citep{Hooker2021}.} This phenomenon has notably entrenched a division between research
in algorithms and software and research in hardware. Hooker posits
that it is therefore necessary to develop methods and techniques to
break down barriers of this kind.

Finally, \citet{Maraninchi2022} considers that the value the discipline
places on the optimisation of software and hardware is contrasted
with `rebound effects' that lead to the expected gains being offset
or even outweighed by increases in usage triggered directly or indirectly
by optimisation. These rebound effects are encouraged by ``a promise
and a deliberate hypothesis'' incorporated in many forms into the
design of computer systems and by which ``resources grow as needed''~\citep[p. 37]{Maraninchi2022}.
The potential incorporation of the concept of ‘limits’ in this field\LyXThinSpace –\LyXThinSpace or,
in other terms, the absence of such promises and hypotheses\LyXThinSpace –\LyXThinSpace is
then described as a ‘paradigm shift’ that, in turn, paves the way
for new research avenues.

\section{The contributions to this special issue}

In this issue, we present contributions on topics drawn from computer
science practice that display characteristics of ‘undone science’,
which means they can be studied through the prism of this concept,
and thereby become genuine examples of ‘undone computer science’.
The overall aim is to analyse the methodological, epistemological,
ethical, social, economic and political issues that these topics shed
light upon. We also aim to bring out problems and questions that may
interest other computer science researchers by encouraging other,
possibly transdisciplinary, questions to be raised that could prove
important for a given field of research but which have tended to be
relegated to informal ‘hallway track’ discussions at conferences,
because they were not considered legitimate according to that research
field’s own standards.

As we have seen, there are already examples of research in the literature
that is moving in this direction. Furthermore, at least one piece
of work is explicitly presented as a case study of ‘undone science’,
namely the article by \citet{Nafus2018}, which considers alternative
ways of exploring databases.\footnote{It is worth noting that \citet{Maraninchi2022} also mentions the
concept of ‘undone science’, even if she does not explicitly situate
her work within this tradition.} Instead of systematically automating the extraction of information
from collected data, this work suggests that a ‘hand made’ analysis
should be carried out by human subjects who are directly concerned
by these data.\footnote{One of the examples discussed in this article concerns data on the
work of people who care for their family members, like the number
of hours of sleep, the ambient noise level in the environment of both
caregivers and those cared for, and so forth.} It goes on to demonstrate that this type of analysis not only enables
more effective ‘cleaning’ of the data in question through the elimination
of irrelevant parts and noise, but also provides insights into the
links between these data and their value to the implied subjects,\footnote{For example, once a caregiver is made aware that a lack of sleep is
affecting their work, they may then decide to hire someone else to
provide night-time care.} independently of the implementation of a computer programme designed
to detect certain critical situations.\footnote{For example, a programme that identifies ``the areas when health
and social systems fail caregivers''\LyXThinSpace –\LyXThinSpace which
could, incidentally, cause stress for them\LyXThinSpace –\LyXThinSpace ``as
opposed to making caregivers individually responsible for their own
stress relief every time that stress is inferred'' \citep[p. 371]{Nafus2018}.}

This special issue also presents three articles that follow a similar
direction to the work of \citet{Nafus2018}, in that they examine
alternative ways of thinking about computing objects and methods.
These papers demonstrate how these alternatives might transform our
conception of computing work from both an epistemological and a social
perspective.

The work of Pierre Saint-Germier, Benjamin Matuszewski and Frédéric
Bevilacqua\footnote{\fullcite{SaintGermier2026}.} focuses on machine
learning and describes alternatives to deep learning methods promoted
widely by Big Tech that are particularly costly in terms of data and
computational resources. Saint-Germier, Matuszewski \& Bevilacqua
are particularly interested in techniques known as ‘shallow learning’
and highlight their epistemological value. Indeed, such methods are
posited as allowing for a form of understanding of learning models
based on the interaction between users and the models themselves (the
authors discuss a machine learning application that enables users
to define correspondences between sounds and gestures).

Sophie Quinton and Jean-Bernard Stefani,\footnote{\fullcite{Quinton2026}.}
for their part, present a research programme focused on the design
and development of computing tools that respect the principle that
Ivan Illich calls `conviviality'~\citep{Illich1973a}. They demonstrate
how, from a normative standpoint, this approach is accompanied by
guiding principles for the design of digital systems, notably by providing
guidelines for integrating computer technologies into a degrowth scenario.

Pierre Depaz\footnote{\fullcite{Depaz2026}.} begins by remarking
that commercial considerations in the field of software engineering
particularly lead to the focus mainly being placed on designing software
whose structure and functionality remain stable when scaled up, particularly
as the volume of input data, the number of concurrent users, or the
number of interconnected computing resources increases. Conversely,
there has not been sufficient research into the transition to a smaller
scale when working locally with limited and specific amounts of data
without pooled computing resources or databases. This perspective
enables Pierre Depaz to consider situations aimed at limiting energy
expenditure, but also gives him the opportunity to bring up epistemological
questions concerning the concept of scale, which he views as closely
linked to the construction of computational artefacts, rather than
as something that is accepted \emph{a priori}.

However, a second meaning can be attached to the idea of ‘undone computer
science’, namely thought about its own epistemological status. Currently,
the focus on the uses of computer science, and on study of the effects
thereof in the economic and social spheres, has left this issue unresolved,
or even unfinished. In other words, is it possible to consider computer
science as a science in its own right and, if so, what kind of science
is it\LyXThinSpace –\LyXThinSpace theoretical or applied? Does computer
science possess its own disciplinary autonomy (and, if so, on what
grounds?) or should it be viewed as a branch of engineering (as is
often the case in the United States, for example, where computer science
departments are frequently integrated into engineering schools)? And
finally, is computer science, as a \emph{science}, completely independent
of the humanities and social sciences?

This special issue also aims to explore these questions, particularly
by examining whether certain philosophical approaches to computer
science tend to neglect certain aspects that are nonetheless central
to its practice\LyXThinSpace –\LyXThinSpace such as its anthropological
and social dimensions\LyXThinSpace –\LyXThinSpace and also essential
to understanding the status of such a discipline. This issue’s fourth
article, by Felienne Hermans,\footnote{\fullcite{Hermans2026}} is
situated within this framework. In this paper, she demonstrates that,
when conceptual reflection on computer science takes place, it is
all too often conceptualised as a science that considers objects or
processes (computation, algorithms, as well as their implementation
and verification) that derive from mathematics, engineering or the
natural sciences, and neglects the human dimension and social interactions,
which do, however, play a crucial role in these practices. 

So, whilst the notion of ‘undone computer science’ may just seem a
form of specialisation or a specific case of ‘undone science’, reflecting
on computer science helps broaden the very concept of ‘undone science’,
beyond the more restricted meaning deriving from studies of social
movements. In this way, thought about computer science as a discipline
encompasses certain parts or subjects that have yet to be studied
in the sense discussed above, or, in other terms, are themselves ‘undone’.
In this sense, this discipline could be considered as possessing a
status that has yet to be fully established and therefore remains
open to change and evolution.

\section{For legitimising computer scientists’\protect \\
reflections about their own discipline}

The fact that computer science does not seem to fit into the traditional
classification of the sciences contributes to making it a \emph{sui
generis} discipline, the boundaries of which are difficult to define
clearly and unambiguously. As Gilles Dowek has noted, from the perspective
of the classification inherited from (particularly logical) positivism,
mathematics deals with \emph{a priori} analytical truth, whereas the
natural sciences deal with \emph{a posteriori} synthetic truths.\footnote{See the interview published in volume 431 of ‘Cahiers de l’INRIA–La
Recherche’ in June 2009: \texttt{\href{https://inria.hal.science/inria-00527531v1}{https://inria.hal.science/inria-00527531v1}}.} However, computer science occupies an entirely unique position, which
Gilles Dowek calls \emph{a posteriori} analytical knowledge, stating
that ``it is both analytical and \emph{a posteriori}.\footnote{Gilles Dowek redefines these terms in a non-Kantian manner. Without
wishing to carry out a detailed analysis of the terminology Dowek
uses, it is possible to posit that he refers to what Kant calls ‘a
priori’ and ‘a posteriori’ as being ‘analytic’ and ‘synthetic’ respectively.
On the other hand, what Dowek calls ``a priori'' and ``a posteriori''
are closer to the Kantian distinction between ``rational knowledge''
and ``historical knowledge'' (Critique of Pure Reason, A836/B864–A837/B865).
In this sense, what Dowek calls ``a posteriori analytical knowledge''
might be close to what Kant refers to as historically acquired a priori
knowledge, that is, knowledge which does not depend on experience
in itself, but can nevertheless be acquired by from empirical data
(\emph{ex datis}). For example, although the truth of the Pythagorean
theorem does not itself depend on sensory experience, a schoolchild
could learn it by heart when they begin school, without being able
to prove it. We would like to thank Baptiste Mélès for drawing our
attention to these distinctions.} It is analytical in the sense that the properties of an algorithm,
for example, are intrinsic (independent of the laws of nature). But
it is also \emph{a posteriori} in the sense that their validation
requires interaction with a physical system (the machine), which is
a form of experimentation in some ways.''\footnote{\emph{Translated from the original French.} This extract can be interpreted
in two ways. On the one hand, this refers to calculability as illustrated
by the halting problem, insofar as only the result of an experiment
generally enables us to find out whether a Turing machine halts, even
though halting is entirely and systematically determined by the machine’s
mathematical definition. On the other hand, it refers back to complexity.
An effective physical implementation of the Turing machine is generally
required to observe whether it halts and then to determine its final
state (if this is even possible within a human timescale).}

However, a machine like a computer is not merely a physical and logical
system (a set of electronic circuits and programmes). It is also a
technological tool that enables certain actions to be carried out
automatically, which, in turn, constitute infrastructures, networks
and ecosystems that then enable forms of interaction and intervention
in the real world. In other terms, as technological instruments, computers
are means by which humans interact with other agents, or which help
shape social and institutional relationships, work and other aspects
of life. 

The unique nature of computer science therefore appears to stem from
its combination of theoretical concepts (the construction of formal
languages, the design of algorithms, the mathematisation of the concept
of information) and technical objects (computers and their constituent
components along with those specific components that enable computers
to be connected to and communicate with each another). However, although
the theoretical model of the Turing machine remains a paradigm in
computer science,\footnote{We use the term `paradigm’ here in the sense of an exemplary past
achievement~\citep[p. 175]{Kuhn1970}.} technical and social developments have, of course, marked computer
science’s brief history. These include, for example, processing speed
(with the miniaturisation of chips from the 1960s onwards), data storage
(which shifted within a few years from punch cards to data centres)
or, finally, access to information (with the emergence of the consumer
internet in the 1990s and the commercialisation of touchscreen handheld
devices that began in the 2010s). The latest significant milestone
in this history came with the recent popularisation of generative
artificial intelligence systems. Technological developments of this
kind have been supported by theoretical work aimed at devising the
most suitable ways of exploiting and optimising the technologies concerned.

This proliferation of information technologies has led to emerging
issues that are linked to the consequences of their widespread adoption
and their control. Issues of this kind have been highlighted in some
of the articles mentioned above. For example, \citet{Green2021} cites
social and political issues while \citet{Maraninchi2022} discusses
environmental concerns. Both of these articles describe the ``crisis
of conscience''~\citep{Green2021} within the profession over the
past decade or so, which is particularly due to the perceived disconnect
between the research carried out and the actual consequences that
have been observed~\citep{Maraninchi2022}. Indeed, although computer
science researchers generally believe their work can contribute to
the development of their discipline and promote the results of their
research for the benefit of society, they are also well particularly
placed to develop critical perspectives on the industry and their
own discipline. This form of tension has been observed to produce
committed, or even militant, stances in defence of certain values
among researchers, along with thought about promising research avenues,
alternatives or explorations of issues that remain marginal, or have
even been totally overlooked. In this respect, approaches of this
kind can be considered as falling under the definition of ‘undone
science’.

In this way, an endogenous dynamic is emerging, in which computer
scientists are choosing to carry out multidisciplinary research into
the practice of computing itself. Work of this kind considers the
uses of computing and the consequences of such uses, along with the
study of the theoretical foundations of certain technical developments
or specific applications of digital technology. And yet, this research
has not always been viewed as legitimate and particularly lacks recognition
within the computer science community itself. Members of this community
tend to consider work of this kind as being more suited to researchers
from the fields of philosophy, sociology, anthropology, or other humanities
disciplines. However, computer scientists are capable of identifying
and expressing other limitations within their discipline and drawing
further conclusions from these, for example by recognising the technical
and theoretical limitations of computing tools. Again, this corresponds
to the concept of ‘undone computer science’ discussed earlier, insofar
as this research shows the discipline of computer science to currently
lack a fully determined status. This is because computer science can
potentially accommodate research by computer scientists into the values,
role and boundaries of the discipline itself, and, for example, exploration
of its links to the humanities.

\section{Conferences on ‘Undone Computer Science’}

To set up a forum for discussion, exchanges and for researchers to
make connections, based on an approach and thinking that are similar
to those discussed above, a group of computer scientists working with
colleagues from the humanities and social sciences began work on the
organisation of a conference that was actually called ‘Undone Computer
Science’ and took place in 2022.\footnote{See \texttt{\href{https://www.undonecs.org/}{https://www.undonecs.org/}}.
The first event was held from February 5th to 7th, 2024, in Nantes
(France), followed by a second conference in Esch-sur-Alzette (Luxembourg)
from March 23rd to 25th, 2026.} One of their aims was to foster the incubation of new ideas and promote
cross-disciplinary collaboration, the end goal being for this thought
and work to lead to published articles.\footnote{The conference was backed up by a separate call for papers that is
open to all, and modelled on the TYPES international conferences.}

The programme committee has invited contributions that involve:\footnote{\texttt{\href{https://www.undonecs.org/2024/en/page/cfp.html}{https://www.undonecs.org/2024/en/page/cfp.html}}.}
\begin{quotation}
``{[}...{]} any discussion of systematic non-production and non-disse\-mination
of knowledge, whether in a specific area or in computer science in
general, either past or present; whether due to limitations of available
methodologies, blind spots of dominant paradigms, institutional and
industrial biases, lack of social representation, or other factors.''
\end{quotation}
\noindent This indeed amounts to an attempt to define the concept
of undone (computer) science that corresponds to the generalisation
outlined above. The quality and quantity of the abstracts submitted
led to the conference being extended by a day.

The seminal article by \citet{Frickel2009} specifically highlighted
the ``analytical potential of undone computer science''. This concept
has indeed proven to be a catalyst for thought and offers an open-ended
line of inquiry capable of addressing multiple aspects of computer
science, including its practice and implications. This concept was
sufficiently precise, general and significant to enable the first
edition of the conference to succeed in bringing together researchers
of different nationalities and from varied backgrounds to discuss
these ideas about the discipline. Around fifty computer scientists
from various subfields attended alongside philosophers, sociologists,
linguists, and researchers in public policy and science and technology
studies, which clearly proved fertile ground for the exchange of ideas
and interdisciplinary dialogue.

Clearly, some may find the term ‘undone computer science’ somewhat
paradoxical when used to define an actual research community. There
is no guarantee that a research programme of this kind\LyXThinSpace –\LyXThinSpace for
example, one that addresses the ethics or epistemology of computer
science in general\LyXThinSpace –\LyXThinSpace might emerge, which
could give structure to such a community and ensure its long-term
viability because, to put it simplistically but self-evidently, research
that has yet to be done stops being non-existent as soon as it is
carried out. Apart from transdisciplinary study of and thought about
computer science, one of the purposes of this community could be to
host research that is destined to slip away from it as soon as a community
of researchers manages to embrace it fully. However, the formation,
growth, and structuring of a community of researchers of this kind
make up an uncertain process that may turn out to be slow in moving
forward. In this scenario, ‘undone computer science’ could be seen
as a form of incubator that promotes the emergence of new computer
science research questions.

\section*{Acknowledgements}

We would like to thank Maël Pégny, Marc Anderson, Sophie Quinton,
Enka Blanchard and \foreignlanguage{vietnamese}{Lê Thành Dũng (Tito)
Nguyễn}, who helped us with our reflections on this subject, and the
many other people involved in the organisation and programme committees
of the ‘Undone Computer Science’ conferences. Alberto Naibo’s work
was partially funded by the ANR GoA project\LyXThinSpace –\LyXThinSpace The
Geometry of Algorithms (ANR-20-CE27-0004).

\printbibliography[heading=bibintoc]

\end{document}